\documentclass[twocolumn,showpacs,preprintnumbers,amsmath,amssymb]{revtex4}
\usepackage{tabularx,graphicx}\begin{document}
\newcommand{\beq}{\begin{equation}}
\newcommand{\eeq}{\end{equation}}
\newcommand{\beqn}{\begin{eqnarray}}
\newcommand{\eeqn}{\end{eqnarray}}
\newcommand{\bmath}{\begin{subequations}}
\newcommand{\emath}{\end{subequations}}
\newcommand{\bk}{\bold{k}}
\newcommand{\bkp}{\bold{k'}}
\newcommand{\bq}{\bold{q}}
\newcommand{\bkb}{\bold{\bar{k}}}
\newcommand{\br}{\bold{r}}
\newcommand{\brp}{\bold{r'}}
\newcommand{\vp}{\varphi}

\title{Hole core in superconductors  and the  origin of the Spin Meissner effect}
\author{J. E. Hirsch }
\address{Department of Physics, University of California, San Diego\\
La Jolla, CA 92093-0319}

\begin{abstract} 
It is  proposed that superconductors possess a hidden `hole core' buried deep in the Fermi sea. 
The proposed hole core is a small region of the Brillouin zone (usually at the center of the zone) where the lowest energy states in the normal state reside.
We propose that in the superconducting state these energy states become {\it singly occupied} with electrons of a definite spin helicity.
In other words, that holes of a definite spin helicity condense from the top to the bottom of the band in the transition to superconductivity, and electrons of that
spin helicity `float' on top of the hole core, thus becoming highly mobile.
The hole core has radius $q_0=1/2\lambda_L$, with $\lambda_L$ the London penetration depth, and the electrons expelled from the hole core 
give an excess negative charge density within a London penetration depth of the real space surface of the superconductor. The hole core
explains the development of a spin current in the transition to superconductivity   (Spin Meissner effect) and the associated negative
charge expulsion from the interior of metals in the transition to superconductivity,   effects we have proposed in earlier work to exist in all
superconductors and to be at the root of the Meissner effect.
  \end{abstract}

\maketitle

\section{Introduction}
 The
Larmor diamagnetic susceptibility for carriers of density $n$ in orbits of radius a is\cite{am}
\beq
\chi_{Larmor}(a)=-\frac{ne^2}{4m_e c^2}a^2
\eeq
It is remarkable that the normal state Landau diamagnetism of metals  is described by this expression for normal state orbits of microscopic radius $k_F^{-1}$:
\beq
\chi_{Landau}=-\frac{1}{3} \mu_B^2 g( \epsilon _F)=-\frac{ne^2}{4m_e c^2}k_F^{-2}=\chi_{Larmor}(k_F^{-1})
\eeq
with $k_F$ the Fermi wavevector, $g(\epsilon_F)=3n/2\epsilon_F$ the free electron density of states, and $\mu_B=e\hbar/2m_ec$ the Bohr magneton.
It is equally remarkable that the perfect diamagnetism of the superconducting state is described by the Larmor formula Eq. (1) for orbits of mesoscopic radius $2\lambda_L$:
\beq
\chi_{London}=-\frac{1}{4\pi}=-\frac{ne^2}{4m_e c^2}(2\lambda_L)^2=\chi_{Larmor}(2\lambda_L)
\eeq
with  the London penetration depth $\lambda_L$ given by the usual form\cite{tinkham}
\beq
\frac{1}{\lambda_L^2}=\frac{4\pi n_s e^2}{m_e c^2}  .
\eeq
and $n_s=n$ the superfluid density. This suggests that (i) the transition to superconductivity involves an $expansion$ of electronic orbits from radius $k_F^{-1}$ to radius $2\lambda_L$, and (ii) that the superconducting ground state
can be understood as being composed of  all electrons ($n_s=n$) residing in orbits of radius $2\lambda_L$. The latter ((ii))
is supported by the observation that the  angular momentum carried by the electrons in the Meissner current 
with velocity $v_s$ in a cylinder of radius $R>>\lambda_L$
and height $h$ can be written in the  two equivalent forms:
\beqn
L_{Meissner}&=&n_s (2\pi R \lambda_L h)\times (m_e v_s R) \nonumber \\
&=& n_s (\pi R^2 h)\times(m_e v_s (2\lambda_L))
\eeqn
The first form is the conventional description of electrons flowing in a surface layer of thickness $\lambda_L$, hence cross-sectional area $2\pi R\lambda_L$, each moving in 
a circle of radius $R$. 
The second form describes all electrons in the bulk, hence cross-sectional area $\pi R^2$, each moving in a mesoscopic orbit of radius $2\lambda_L$.
 
The proposition that  the transition to superconductivity involves an expansion of electronic orbits from radius $k_F^{-1}$ to radius $2\lambda_L$ provides a $dynamical$
explanation of the Meissner effect through the Lorentz force acting on radially outgoing electrons\cite{sm,copses}, an effect  which does not exist in the conventional theory of superconductivity. We have argued that the conventional theory is 
untenable\cite{bcs}  because it cannot provide a dynamical explanation of the Meissner effect\cite{emf} and because it leads to non-conservation of angular momentum\cite{missing}.

We showed in Ref.\cite{sm}   that in the absence of an applied magnetic field 
this orbit expansion in the presence of the ionic electric field will generate through the spin-orbit interaction
 a spin current near the surface of the superconductor, with  carrier velocity given by
\beq
\bold{v}_\sigma^0=-\frac{\hbar} {4 m_e \lambda_L} \bold{\sigma} \times \bold{\hat{n}}
\eeq
with $\bold{\hat{n}}$ the normal direction to the surface of the superconductor (Spin Meissner effect). The spin current originates in the superposition of  real space   orbits of radius $2\lambda_L$ for electrons moving with
speed Eq. (6), each electron thus carrying orbital angular momentum $\hbar/2$\cite{sm}.

Electronic orbit expansion and the associated radially outgoing electron flow lead to a non-homogeneous charge distribution in the superconducting state, with extra
negative charge near the surface, as proposed within the theory of hole superconductivity\cite{chargeexp1,chargeexp2}. New electrodynamic equations
describe this scenario\cite{edyn,electrospin}, that yield a definite relation between the magnitude of the expelled
negative charge and resulting electric field in the interior of superconductors, and the velocity of carriers in the spin current Eq. (6) (Ref.\cite{electrospin}, Sect. 4).
These equations lead to the surprising prediction that the 
density of carriers of given spin orientation near the surface will change as a function of the magnitude of the applied magnetic field and resulting 
charge current (Ref.\cite{electrospin}, Eq.(26c) and\cite{japan2}), leading to a
non-zero Knight shift in the superconducting state, consistent with   experiments\cite{knight}.

In addition, we have proposed in earlier work that the gap function in superconductors has a gap slope of universal sign
 (Ref.\cite{gapslope}, Eq. (19) and Fig.7)
 (which manifests itself experimentally in an observed tunneling
asymmetry of universal sign\cite{tunnasym}), which gives rise to a shift in the chemical potential in going from the normal to the superconducting state of magnitude determined by the slope of the gap function\cite{photo}, and that   superconductivity is driven by kinetic energy lowering\cite{kinen1,kinen2} and `undressing'  of 
carriers\cite{giantatom,hawai}  from the electron-electron as
well as from the electron-ion interaction\cite{holeelec2}.

In this paper we show that all  these phenomena can be understood as arising from the existence of a hidden `hole core' buried deep  inside the Fermi sea of metals in the superconducting state.

\section{Hole core}
 
 If an electric field exists in the interior of superconductors\cite{chargeexp2}, it is only natural to expect that the Rashba spin-orbit interaction\cite{rashba} will play an important role.
We assume   that in the normal state electronic energies are given by the free electron dispersion relation
$\epsilon_k= \hbar^2 k^2/2m_e$
and that in the superconducting state spin-split bands develop, with dispersion
\beq
\epsilon_{\bk \sigma}=\frac{\hbar^2 k^2}{2m_e} -\frac{\hbar^2}{2m_e}kq_0 \vec{\sigma}\cdot(\hat{k}\times\hat{n}) .
\eeq
where $\hat{n}$ is the outward-pointing normal to the surface of the superconductor and
\beq
q_0=\frac{1}{2\lambda_L}  .
\eeq
The energy dispersion Eq. (7) can be understood as arising from the Dirac spin-orbit interaction\cite{so}
  \beq
H_{s.o.}=-\frac{e\hbar}{4m_e^2 c^2}\vec{\sigma}\cdot(\bold{E}\times\bold{p})
\eeq
with $\bold{p}=\hbar\bk$ and electric field
\beq
E=2\pi\rho_i r
\eeq  
pointing outward (parallel to $\hat{n}$). The electric field Eq. (10) is the radial electric field generated at the surface of a cylinder of radius $r=2\lambda_L$ by the
ionic charge density $\rho_i=|e|n_s$ that compensates the superfluid electronic charge density $en_s$ ($n_s=$ superfluid density).
 The equality between Eqs. (9), (10) and the second term in Eq. (7) follows from using the usual relation for the London penetration depth Eq. (4).
In this interpretation, the spin-orbit interaction can be understood
as arising from the $unscreened$ positive charge density $|e|n_s$ acting on the macroscopic superfluid wave function. Instead, for a single electron 
in the presence of the electric field
\beq
E_m\hat{n}=-\frac{\hbar c}{4e\lambda_L^2} \hat{n}
\eeq
 that arises from charge expulsion\cite{electrospin}, the Dirac spin-orbit interaction gives    the second term in Eq. (7) 
 reduced by the factor
 $r_q q_0$, with $r_q=\hbar/2m_e c$ the `quantum electron radius'\cite{holeelec4}.

 The dispersion relation Eq. (7) gives rise to two 'Rashba-like' bands, given by
 
 \beq
 \epsilon_k^1=\frac{\hbar^2k^2}{2m_e}-\frac{\hbar^2kq_0}{2m_e}=\frac{\hbar^2}{2m_e}(k-\frac{q_0}{2})^2-\frac{\hbar^2 q_0^2}{8m_e}
 \eeq
  \beq
 \epsilon_k^2=\frac{\hbar^2k^2}{2m_e}+\frac{\hbar^2kq_0}{2m_e}=\frac{\hbar^2}{2m_e}(k+\frac{q_0}{2})^2-\frac{\hbar^2 q_0^2}{8m_e}
 \eeq
 corresponding to spin orientations $\vec{\sigma}\cdot(\hat{k}\times\hat{n})=\pm1$. The bands are shown in Figure 1. There are negative
 energy states in band 1 in the range $0<k<q_0$. We postulate that as the system goes superconducting
 the spin-degenerate band  splits into the two bands Eqs. (12),  (13) $and$ simultaneously the negative energy states in band
 1 fill up with holes, driven by the Coulomb interaction\cite{holeelec4}. In other words, electrons in these negative energy states get ejected from the bulk of the superconductor
 towards the surface. The occupation of the bands in the Brillouin zone is shown in Figure 2. The gap between energy states in both bands for the same wavevector is
 \beq
 \Delta_k=\frac{\epsilon_k^1-\epsilon_k^2}{2}=\frac{\hbar^2 k q_0}{2 m_e}
 \eeq
 as obtained in Ref. \cite{copses} using different arguments.

    \begin{figure}
\resizebox{8.5cm}{!}{\includegraphics[width=7cm]{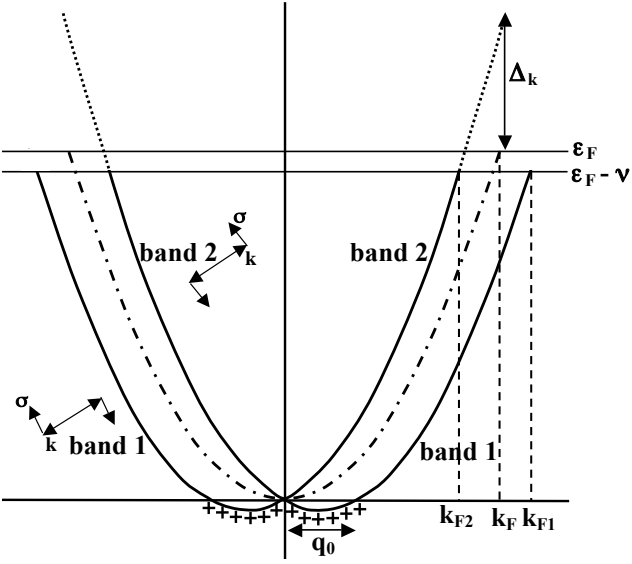}}
  \caption {Spin-split bands below $T_c$ given by Eqs. (12), (13). $k_F$  and $\epsilon_F$ are the Fermi wavevector  and Fermi energy
  (chemical potential) in the normal state with energy dispersion relation  indicated by the dot-dashed line. 
  In the superconducting state the Fermi energy shifts down by $\nu$ and the Fermi wavevectors for the two
  Rashba bands  are $k_{F1}$ and $k_{F2}$.
  The negative energy states in band 1  in the region $k<q_0$ are filled with holes as indicated by the  $+$ signs.
 Note that in the region $k_{F2}<k<k_{F1}$  only electrons of helicity $\vec{\sigma}\cdot(\hat{k}\times\hat{z})=1$ exist.}
\end{figure}

 \begin{figure}
\resizebox{8.5cm}{!}{\includegraphics[width=7cm]{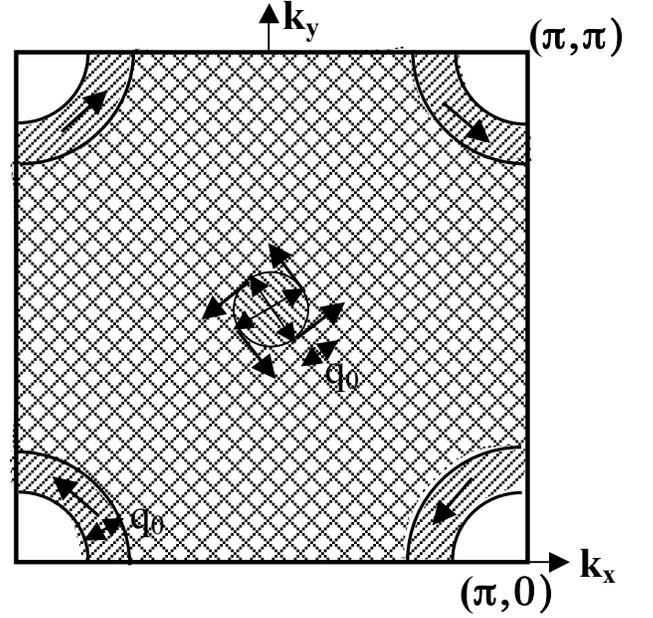}}
  \caption {Brillouin zone occupation for the spin-split bands of Fig. 1.  The   $\hat{n}$ axis points out of the paper.
  A circle of radius $q_0=1/2\lambda_L$ in the $(k_x,k_y)$ plane centered at $(0,0)$ is
  singly occupied by electrons of helicity $\vec{\sigma}\cdot (\hat{k}\times\hat{n})=-1$. The arrows near the center indicate the direction of 
  $(k_x,k_y)$ (radial arrow) and of the spin of the corresponding electron (attached tangential arrow). There is an excess of electrons 
    of helicity $\vec{\sigma}\cdot (\hat{k}\times\hat{n})=1$ at the  Fermi surface in the antibonding regions close to $(k_x,k_y)=(\pm \pi, \pm \pi)$, (their
  spin orientation is indicated by the arrows). The resulting regions
  with singly-occupied electrons are denoted by the hatched areas, with $45^\circ$ hatching   (-$45^\circ$ hatching) in the
  antibonding  (bonding) regions respectively.  
  The region hatched by the square pattern is doubly occupied with electrons, and the white regions near $(k_x,k_y)= (\pm \pi, \pm \pi)$ are
  doubly occupied with holes.}
\end{figure}
 
 The carrier density for a two-dimensional free electron band structure is given by
 $n_s=k_F^2/2\pi$, 
  the Fermi energy by
$\epsilon_F=\hbar^2k_F^2/2m_e$, 
  the total energy (per unit area) by
  \beq
  E=\frac{1}{\pi}\frac{\hbar^2}{2m_e}\frac{k_F^4}{4} 
    \eeq
  and the energy per particle by $
  \epsilon=\epsilon_F/2= \hbar^2 k_F^2/4m_e$. 
For the spin-split bands, assuming no change in the density of particles, the new Fermi wavevectors are
  \beq
  k_{F1}=\sqrt{k_F^2-q_0^2/4}+q_0/2
  \eeq
    \beq
  k_{F2}=\sqrt{k_F^2-q_0^2/4}-q_0/2
  \eeq
  and the new Fermi energy is
  \beq
  \epsilon_{k_{F1}}=  \epsilon_{k_{F2}}=\epsilon_F-\frac{\hbar^2 q_0^2}{4m_e} \equiv \epsilon_F- \nu
  \eeq
Thus, the Fermi energy (or chemical potential) is lowered by $\nu$ in the transition to superconductivity. This change in chemical potential as
the system goes superconducting was derived within the theory of hole superconductivity as originating in the slope of the gap function\cite{photo} 
\beq
\nu=\frac{1}{2}\frac{\partial}{\partial \epsilon_k}(\Delta_k)^2
\eeq
and coincides with the result Eq. (18) derived here from the 
 condition of constant density of particles.

The energy of electrons  in the spin-split bands (per unit area) is given by
\beq
E_i=\frac{1}{2\pi} \frac{\hbar^2}{2m_e} (\frac{k_{Fi}^4}{4}-\sigma_i\frac{q_0 k_{Fi}^3}{3})
\eeq
with $\sigma_1=1, \sigma_2=-1$. Thus, electrons in band 1 lower their kinetic energy upon spin splitting and those in band 2 raise their
kinetic energy. The net change in energy per particle  is 
\beq
\Delta \epsilon=\frac{1}{n_s}(E-E_1-E_2)=  \frac{\hbar^2 q_0^2}{4m_e}=\nu
\eeq
in agreement with the result obtained  in Ref.\cite{copses} using entirely different arguments. Eq. (21) is the condensation energy per particle
in the superconducting state.  Thus, the condensation energy in this model can be interpreted
as arising from kinetic energy lowering (in band 1), in agreement with previous considerations\cite{kinen1,kinen2}.

The number density in each band is given by
\beq
n_1=\frac{n_s}{2}+\frac{q_0k_F}{4\pi}-\frac{q_0^2}{4\pi}
\eeq
\beq
n_2=\frac{n_s}{2}-\frac{q_0k_F}{4\pi} 
\eeq
The last term in Eq. (22) is the missing density of carriers in the negative energy states, which we argue is expelled to the surface of the
superconductor
\beq
n_{exp}=\frac{q_0^2}{4\pi}
\eeq

The speed of carriers in the spin-split bands is given by $v_k^i=(1/\hbar)\partial\epsilon_k^i/\partial k$, hence
\beq
v_k^1=\frac{\hbar}{m_e}(k-\frac{q_0}{2})
\eeq
\beq
v_k^2=\frac{\hbar}{m_e}(k+\frac{q_0}{2})
\eeq
Therefore, the speed of the carriers at the large Fermi surface is the same for both bands ($v_{kF1}^1=v_{kF2}^2$).
However, at the small Fermi surface ($k=q_0$) we have
\beq
v_k^1=\frac{\hbar q_0}{2m_e}=v_\sigma^0
\eeq
Thus, we can interpret Eq. (27) as the spin current speed of the condensate as a whole. In a cylindrical geometry, the carriers near
the surface with spin orientation parallel or antiparallel to the cylinder axis will have a `drift' velocity given by $\pm v_\sigma^0$ parallel
to the surface and perpendicular to the cylinder axis, as proposed in our paper on the Spin Meissner effect based on entirely
different arguments\cite{sm}.

 \begin{figure}
\resizebox{8.5cm}{!}{\includegraphics[width=7cm]{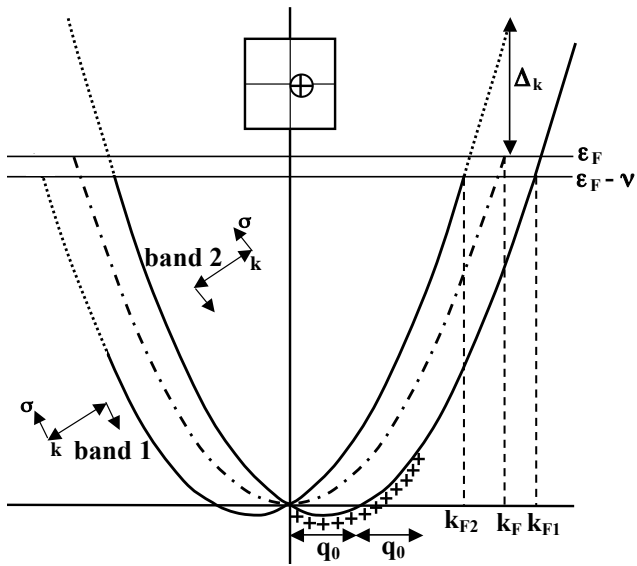}}
  \caption {Spin-split bands in the presence of a charge current.  
  The occupation in band 1 shifts (the occupied states are depicted by the solid line) while that in band 2 remain unchanged. 
  The situation depicted corresponds to the critical current where the wavevector shift is $q_0$. The position of the hole core in the Brillouin zone is shown at the top.}
\end{figure}

When a charge current circulates, we argue that it is only the electrons in band 1 that shift their occupation.
This is based on the fact that experiments on rotating superconductors\cite{rotating1,rotating2} can be interpreted as indicating\cite{ehasym} that the 
superconducting current is carried by completely `undressed' electrons, suggesting that   a shift in the small Fermi surface describing long wavelength free-electron-like carriers
is involved. Hence we assume that the current density is given by
\beq
J_s=e\frac{n_s}{2}v_s^1=-\frac{n_s e^2}{m_e c} \lambda_L B
\eeq
involving only half the superfluid density ($n_s/2$) corresponding to the carriers in band 1, moving with superfluid velocity $v_s^1$. The second equality in Eq. (28) follows from London's 
equation. Thus Eq. (28) yields for the drift speed of the carriers in band 1
\beq
v_s^1=\frac{2e}{m_e c}\lambda_L B
\eeq
corresponding to a shift in their wavevector
\beq
\Delta q=\frac{m_e}{\hbar}v_s^1=\frac{2e}{\hbar c}\lambda_L B
\eeq
When the wavevector shift reaches the value $q_0$ the small Fermi surface crosses the origin and it is reasonable to expect that this corresponds to the critical value of
magnetic field where superconductivity will be destroyed. Eq. (30) then yields for the critical field (when $\Delta q=q_0$)
\beq
B_c=-\frac{\hbar c}{4e\lambda_L^2}
\eeq
which is essentially $H_{c1}$, the lower critical field of type II superconductors\cite{tinkham}. The result Eq. (31) coincides 
with the result obtained analyzing the Spin Meissner effect based on entirely different arguments\cite{sm}.
Figure 3 shows the occupation of the bands  in the presence of a charge current.

 \begin{figure}
\resizebox{8.5cm}{!}{\includegraphics[width=7cm]{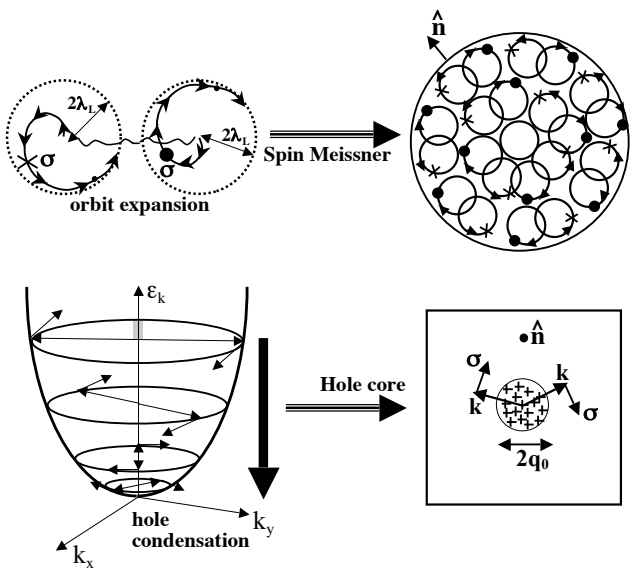}}
  \caption {In real space, orbits expand from a microscopic length to radius $2\lambda_L$ (upper left). The upper right picture shows the electronic orbits 
  in a cylinder viewed along the axis, electrons with spin out of (into) the
  paper circulate clockwise (counterclockwise)
  with speed $v_\sigma^0$ (Eq. (1)) . The lower left panel shows pairs of holes of helicity $\vec{\sigma}\cdot(\hat{k}\times\hat{n})=1$ dropping from the top of band 1 to the
  bottom to form the hole core, shown in the lower right panel, with  directions for $\bk$ and $\vec{\sigma}$ for holes
  in the hole core  indicated schematically. }
\end{figure}

In connection with the Spin Meissner effect we showed\cite{sm} that the spin current with speed Eq. (6) develops when carriers expand their
orbits from a microscopic length to radius $2\lambda_L$ and are deflected azimuthally by the spin-orbit interaction. 
This corresponds in the current scenario to the dropping of holes in band 1 from the top to the bottom to form the hole core. The two representations of the same physics are shown in Fig. 4.
Furthermore we showed in refs. \cite{sm,copses} that the orbit expansion in the presence of a magnetic field gives rise to the charge current needed for the 
Meissner effect, and argued that the orbit expansion is driven by kinetic energy lowering. Clearly, in the current scenario this corresponds
to the carriers in band 1 developing a charge current as the holes condense from the top of the band to the hole core in the presence of a magnetic field,
with the kinetic energy of carriers in band 1 being lowered as described by Eq. (21).

The scenario depicted in Fig. 3 also shows that when a charge current circulates the magnetization of the system will change. 
Assuming that the ejected electrons from the states indicated by `{\bf +}' in Fig. 3 represent the excess electrons giving rise to the negative charge density
$\rho_-$ predicted to exist near the surface by the modified electrodynamic equations\cite{edyn}, Fig. 3 shows that as the critical magnetic field is
approached the excess electrons moving in direction opposite to the charge current all have the same spin orientation, and the excess electron density of
opposite spin orientation has been depleted to zero (since all the holes have moved to the branch on the right in Fig. 3). This is precisely the situation
predicted by the electrodynamic analysis of ref.\cite{electrospin} using entirely different arguments. We found in that work that the expelled charge density 
for spin $\vec{\sigma}$ is
\beq
\rho_\sigma=\frac{en_s}{2}\frac{v_\sigma^0}{c}+\frac{1}{8\pi \lambda_L} \vec{\sigma}\cdot \vec{B}
\eeq
so that the expelled charge density for $\vec{\sigma}$ parallel to $\vec{B}$ goes to zero at the critical field Eq. (31), in agreement with Fig. 3. The variation of spin density with magnetic field
indicates that the superconducting state has a triplet component and will give rise to a non-zero Knight shift at zero temperature\cite{japan2}.
It is remarkable that the electrodynamic analysis of Ref.\cite{electrospin} predicted this `Edelstein effect'\cite{edelstein}, which we can now
understand as originating in the locking of spin and momentum variables in the Rashba-like bands.

Finally we would like to show that the expelled electrons from the negative energy states of band 1 correspond to the excess negative charge near the surface derived
from the electrodynamic equations, given by\cite{electrospin}
\beq
\rho_-=en_s\frac{v_\sigma^0}{c} .
\eeq
This excess charge is distributed over a surface layer of thickness $\lambda_L$, hence the `surface charge density' $\Sigma$  is
\beq
\Sigma=\rho_- \lambda_L=en_s\frac{\hbar}{4m_e c}
\eeq
The charge content of the hole core (Eq. 24) is     
\beq
n_{exp}=\frac{q_0^2}{4\pi}=n_s\frac{\hbar}{4m_e c}\frac{e^2}{\hbar c}=(\Sigma/e)\times \alpha
\eeq
with $\alpha=e^2/\hbar c=1/137$ the fine structure constant. Thus, Eq. (35) implies that the expelled negative charge density $\rho_-$ originates in the hole core of $137$ lattice
planes, which itself is of the order of $\lambda_L$, as first suggested by Slater\cite{slater}.

Alternatively we can argue: if $\Delta z$ is the spacing between lattice  planes parallel to the surface, the three-dimensional charge density resulting from Eq. (24) is 
 \beq
 n_{exp}^{3d}=\frac{q_0^2}{4\pi \Delta z}
 \eeq
and setting $n_{exp}^{3d}=\rho_-/e$ yields $\Delta z=\alpha\lambda_L$. Taking for the spacing between lattice planes $\Delta z=2a_0$, with $a_0=\hbar^2/m_e e^2$ the Bohr radius yields
 \beq
 n_s=\frac{1}{2\pi}\frac{1}{(2a_0)^3}
 \eeq
 corresponding, for a simple cubic lattice of lattice constant $a=2a_0$, to a band filling $1/2\pi=0.16$ holes per atom, an almost filled band, consistent with the assumptions of the theory
 of hole superconductivity.

\section{Discussion}
Within the theory of hole superconductivity, negative charge is expelled from the interior to the surface of a metal in the transition to superconductivity.
A spin current exists in the superconductor near the surface, and the superconducting state can be understood as a superposition of electronic orbits of radius $2\lambda_L$.
Superconductivity is driven by kinetic energy lowering, the chemical potential drops in the transition to superconductivity by an amount $\nu$ related to 
the slope of the gap function versus energy which has universal sign, and $\nu$ also gives the condensation energy per carrier. The spin density near the 
surface varies with the applied magnetic field and associated superfluid velocity. Carriers `undress' from the electron-ion interaction, expand their wavelength
and become free-electron-like as the system becomes superconducting.

All these properties were deduced in earlier work based on different arguments. In this paper we have shown that they can all be understood within a scenario where
a `hole core' of radius $q_0=1/2\lambda_L$ with hole carriers of helicity $\vec{\sigma}\cdot(\hat{k}\times\hat{n})=1$ develops at the 
bottom of the electronic energy band.
The direction $\hat{n}$ is the direction of the interior electric field pointing towards the closest surface. The electric field arises due to negative charge being expelled 
from the hole core states to the
surface of the superconductor. Thus, the scenario proposed here is consistent with and supports the Spin Meissner effect scenario\cite{sm} and its relation with negative charge expulsion deduced from
the electrodynamic equations\cite{electrospin}.  The `hole core'  is the k-space representation of
this physics.

Note that in the scenario proposed here, `Cooper pairs' formed by time-reversed carriers, $(\bk\uparrow,-\bk\downarrow)$, are essential, with
the spin directions  $\uparrow$ and $\downarrow$ corresponding
to $\vec{\sigma}\cdot(\hat{k}\times\hat{n})=\pm1$ rather than being fixed in space.  Note also that the number of holes in the hole core is only
$\sim10^{-6} n_s$, so that the energy cost in promoting those electrons to the Fermi energy is only of  order $\mu eV$ per superfluid electron.

We believe that a full description of the superconducting state of matter has to include the physical elements discussed here
(which are not part of conventional BCS theory) and that this physics should be
experimentally verifiable. In a planar geometry, the direction of the electric field in the interior
and the direction of the majority electron spin for given $\bk$ direction 
is shown in Fig. 5. The k-space scenario shown in Figure 2 should be testable by  high resolution spin-resolved angle-resolved photoemission\cite{dessau,spinres},
keeping in mind that depending on the particular band structure the location of the spin-split regions could be different from those shown in Figure 2.

 \begin{figure}
\resizebox{8.5cm}{!}{\includegraphics[width=7cm]{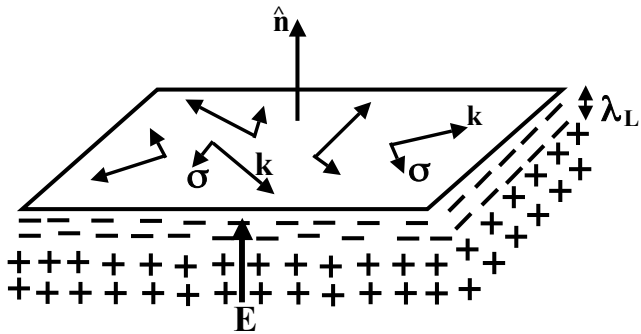}}
  \caption {The plane shown is the boundary of the superconductor. An electric field in the interior of the
  superconductor points in the $\hat{n}$ direction. There is excess negative charge  within distance
  $\lambda_L$ from the surface of the superconductor that was expelled from its interior that screens the
  interior electric field, and there is a spin current near
  the surface. The long and short perpendicular arrows denote the 
  direction of the momentum and the dominant carrier's spin. }
\end{figure}

 \end{document}